# Acoustic Probing for Estimating the Storage Time and Firmness of Tomatoes and Mandarin Oranges

Short Title: Acoustic Probing for Estimating Storage Time of Fruits

Hidetomo Kataoka[1],   Takashi Ijiri[2,*],   Kohei Matsumura[1],   Jeremy White[1],   Akira Hirabayashi[1]

1. Graduate School of Information Science and Engineering, Ritsumeikan University, Kusatsu, Shiga, Japan
2. College of Engineering, Shibaura Institute of Technology, Toyosu, Tokyo, Japan

**\*** Corresponding author:

E-Mail: takashi.ijiri80@gmail.com

# Acoustic Probing for Estimating the Storage Time and Firmness of Tomatoes and Mandarin Oranges

Hidetomo Kataoka[1], Takashi Ijiri[2], Kohei Matsumura[1], Jeremy White[1], Akira Hirabayashi[1]

1. Ritsumeikan University,    2. Shibaura Institute of Technology

**Abstract**

This paper introduces an acoustic probing technique to estimate the storage time and firmness of fruits; we emit an acoustic signal to fruit from a small speaker and capture the reflected signal with a tiny microphone. We collect reflected signals for fruits with various storage times and firmness conditions, using them to train regressors for estimation. To evaluate the feasibility of our acoustic probing, we performed experiments; we prepared 162 tomatoes and 153 mandarin oranges, collected their reflected signals using our developed device and measured their firmness with a fruit firmness tester, for a period of 35 days for tomatoes and 60 days for mandarin oranges. We performed cross validation by using this data set. The average estimation errors of storage time and firmness for tomatoes were 0.89 days and 9.47 g/mm$^2$. Those for mandarin oranges were 1.67 days and 15.67 g/mm$^2$. The estimation of storage time was sufficiently accurate for casual users to select fruits in their favorite condition at home. In the experiments, we tested four different acoustic probes and found that sweep signals provide highly accurate estimation results.

## 1. Introduction

Each type of fruit has their best maturity rate. Ways of eating fruits, such as raw or cooked, is selected based on their maturity rate. Storage time of fruits is important for consumers, since it is strongly related to the maturing rate and deterioration. However, it is often difficult for consumers to know the storage times of fruits when buying or eating them, because expiry dates or harvesting dates of fruits are rarely displayed in supermarkets. One way to know the maturing rate of a fruit is examining its firmness by touching it. However, touching fruits may cause damage on them and sometimes it is not allowed in supermarkets. Besides, consumer without expertise cannot examine the maturity rate using this approach.

Various methods have been developed to measure the quality of fruits. They can be roughly classified into three groups based on the measurement approaches [1], such as optical [2-7], electrochemical [8], and mechanical [9-17] approaches. The optical approach measures optical features. In particular, near-infrared spectroscopy (NIRS) measurement for nutrient content estimation is popularly used. Electrochemical approach measures the amount of gas released from fruits. The amount of ethylene gas is commonly used to estimate maturity. The mechanical approach directly contacts to a target and measures its physical features such as firmness. We survey them in Section 2 in detail. These existing methods, however, are mainly developed and used by producers and distributors of fruits, requiring professional, and often expensive and space consuming devices. It is still difficult for consumers to know the storage time and firmness of fruits easily.

Our goal is to present an easy-to-use technique that allows consumers to estimate storage time and firmness of fruits. The key idea is to utilize an acoustic probe; we provide an acoustic signal to a fruit by attaching a small speaker onto its surface and capture the reflected signal by a microphone. We collect reflected signals from target fruits with various conditions and extract feature vectors from the reflected signals. We then train regressors with the feature vectors to estimate the storage time and firmness of fruits.

To evaluate the accuracy and feasibility of our acoustic probing technique, we performed data collection and cross validation. We collected reflected signals and firmness values of 162 tomatoes for a period of 35 days and 153 mandarin oranges for 60 days. We then trained regressors with a part of the collected data and examined the accuracy of the regressors by using the rest of the data. As a result, we found sufficiently accurate estimation results for the storage time; the average estimation errors of storage time and firmness for tomatoes were 0.89 days and 9.47 g/mm$^2$, and those for mandarin oranges were 1.67 days and 15.67 g/mm$^2$. During the experiments, we tested two feature vectors, spectrum and Mel-Frequency-Cepstrum-Coefficients (MFCCs) and two regressors, support vector regression (SVR) and gradient boosting regression (GBR). In addition to this, to select appropriate probes, we tested four different probing signals, such as single-frequency, multi-frequency, linear-sweep and exponential sweep signals. We found that the spectrum feature and sweeps signals provided highly accurate estimation results.

The contributions of this paper are listed as follows:

- We present an acoustic probing technique to estimate the storage time and firmness of fruits.

- We tested four types of probe signals and found that the probes containing various spectrum result in accurate estimation.

- We achieved a less invasive and easy-to-use technique that only captures acoustic signals and requires common sensors.

## 2. Related Work

Many methods have been developed for measuring attributes related to quality of fruits and vegetables. The quality includes visual attributes such as color, shape and texture, and non-visual attributes such as maturity, firmness, nutritional components and internal defects. According to Abbott et al. [1], existing methods can be classified into three groups based on their measurement approaches; optical, electrochemical, and mechanical approaches. In particular, our technique uses acoustic signals and is classified into the mechanical approach. After reviewing each of the three below, we also survey studies utilize acoustic sensing.

## 2.1 Optical Approach

The optical approach measures optical signals. It includes the image-based approach [2, 3], hyper spectral imaging (HSI) [4, 5], near infrared spectroscopy (NIRS) [6, 7], and refraction meters. Some researchers use photographs to estimate ripeness of watermelons [2] or to measure the volume of oranges [3]. However, these image-based approaches require specific conditions to take photographs. HSI is adopted to estimate quality of various fruits, such as strawberries [4]. NIRS is also used to estimate the quality of fruits, such as apples [6]. Surveys by Zhan et al. [5] and Nicolai et al. [7] provide more examples for HIS and NIRS, respectively. However, the both HSI and NIRS require specific devices and are difficult for casual users to access. The Refraction meter (or Brix meter) measures the refractive index of fruit juice. Although this measurement is achieved by a small and cheap device, the measurement process is invasive since it requires fruit juice.

## 2.2 Electrochemical Approach

Electrochemical approach measures the amount of gas released from a sample. Ethylene gas measurement is popularly used to estimate maturing rate of fruits [8]. Since this method achieves a non-destructive and accurate estimation for maturing rate and deterioration, it is widely used by producers and distributors. However, it requires expensive and large devices.

## 2.3 Mechanical Approach

The mechanical approach measures firmness or impact-response of a sample by using a device physically contacting a sample. Firmness of fruits is measured by pushing fruit surfaces with a cylindrical probe and recording deformation or penetration forces [9, 10]. The firmness tester is usually compact and not expensive; however, it pushes and pierces the sample. Some researchers measure impact-response of fruits [11, 12] or cheese [13]; they physically hit a sample and capture the response signal by a microphone to estimate their conditions. For instance, the existence of hollow heart in a watermelon [11] and firmness of peaches [12] were estimated using this approach. The same approach is also adopted for estimating maturing rate of Manchego cheese [13]. Although these methods measure the maturing rate and firmness of samples, they require custom devices to provide an impact.

Acoustic impulse-response measurement is adopted for fruits quality estimation. Since this approach provides an impulse by using a microphone contacting to a sample, we would like to categorize it into the group of mechanical approach. Muramatsu et al. [14] provide a pulsed sound to a kiwifruit and measure the elapsed time of the transmitted signal to estimate firmness. Similarly, Schotte et al. [15] and Belie et al. [16] apply acoustic impulse-response measurements to firmness estimation of tomatoes and pears, respectively. However, these methods are limited to impulse signal probing and to firmness estimation. To explore the broader applicability of the acoustic probing, our research group apply a sweep signal to estimate storage time of tomatoes [17]. This preliminary study, however, has limitations in the number of samples, the variety of samples and the variation of probing signals. To further confirm the capability of the acoustic probing, this paper presents experiments with approximately 160 samples and with two different fruits. In addition, we tested four different probing signals.

## 2.4 Acoustic Sensing

The acoustic proving is used for a different purpose. SoQr system [18], which strongly inspires our work, estimates content level inside containers, such as plastic water bottles and potato chips packages. They provide sweep signals to a bottle or packages and captures transmitted signals to train a classifier. We have extended this idea to include freshness and firmness of fruits.

In the realm of computer science, especially in the area of human-computer interaction (HCI) and ubiquitous computing, scholars utilize acoustic sensing techniques, for example, acoustic barcodes for object identification [19], swept sound [20], ultrasound signal [21], air pulses sensing [22] and tine-strike sound [23] for detecting user inputs, ultrasound propagation for detecting gesture on the arm [24] and acoustic signature for identifying the location of a device [25]. We provide swept-frequency acoustic signals to fruits and measure the reflected signal to estimate the storage time and firmness. The interaction between foods and humans has become a target for interaction design and HCI [26, 27]. Our research will allow a way for developing a method for estimating food freshness and will be a steppingstone to advance of the field of HCI.

## 3. Method

### 3.1 Overview

Our goal is to develop a less-invasive and easy-to-use estimation technique for storage time and firmness of fruits by using common and cheap sensors. The key idea is to utilize acoustic probing. As illustrated in Fig 1, our technique consists of two phases: training and estimation. In the training phase, we collect sample fruits with various conditions. We provide an acoustic signal to a sample and capture a reflected signal by using a device consists of a small speaker and microphone (Section 3.2). We also record storage time and firmness values of the samples. For each captured signal, we extract feature vector by clipping it and converting it to a spectrum or MFCCs (Section 3.3). We finally train regressors with the collected feature vectors, storage times and firmness values of samples. In the estimation phase, given a target fruit, we measure its reflected signal and extract a feature vector similarly to the training phase and estimate a storage time and firmness by using trained regressors.

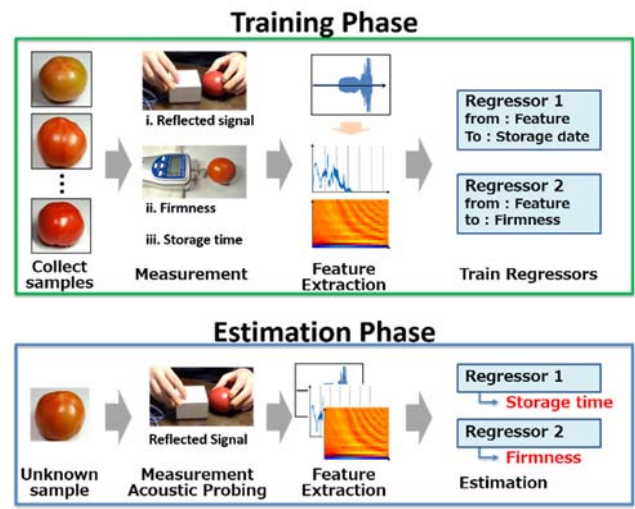

Fig 1. Overview of our technique. In the training phase, we collect sample fruits with various conditions, collecting training data sets from them, extracting feature vectors from captured reflected signals, and training regressors. In the estimation phase, we capture a reflected signal of a novel sample, extract a feature vector and estimate its storage time and firmness by using the trained regressors.

### 3.2 Acoustic probing

Our probing device is shown in Fig 2a. Its size is 65 mm × 65 mm × 45 mm. It contains a small speaker and microphone; we used a canal type earphone, *SONY MDR-EX150*, as a speaker, and a silicon microphone, *Knowles SPU0414HR5H-SB*, with an amplifier. During the capturing phase, we placed the device at the side of a sample fruit so that the earphone gently contacted to the sample and the distance from the microphone to the sample surface is about 5 mm (Fig 2b). In this condition, we provide an acoustic probe from the speaker and capture a reflected signal using the microphone. The sampling rate was 48 kHz.

Since our target users are consumers, a measurement device needed to consist of common, cheap, and small sensors. Our probing requires only a speaker and microphone, and it is thus possible to construct a probing device with low cost and small size. We only provide and capture acoustic signals, and thus the damage caused by the measurement is very small.

Because acoustic probes that will realize high estimation accuracy are unknown, this study examines the following four types of signals (Fig 2c-g). *Single-frequency* signal consists of a 5000 Hz sine wave and *Multi-frequency* signal consists of 3500 Hz, 5000 Hz, and 6500 Hz sine waves. *Linear-sweep* signal linearly sweeps from 100 Hz to 10000 Hz and *exponential-sweep* signal exponentially sweeps the same frequency range as the linear-sweep. We selected them based on [17, 18]. Each of them is one second long. For efficient data collection in the training phase, we joined the four probes in the order of single-frequency, exponential-sweep, linear-sweep, and multiple-frequency within a single sound file (Fig 2g) and captured reflected signals of the four probes at a time.

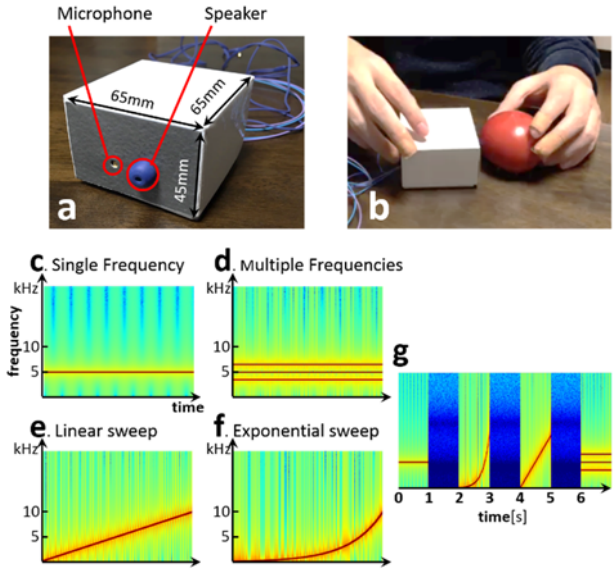

Fig 2. Device for our acoustic probing. We measured reflected acoustic signals by contacting our probing device (a) to a sample (b). Four candidate acoustic probes (c-f) are joined into a file for efficient data collection (g).

### 3.3 Feature vector extraction

To train regressors, we extracted a feature vector from each reflected signal. The captured raw signal includes intervals in which probes are not provided (Fig. 3ab). We then clipped the raw signal into one second so as to extract a part corresponding to a probe (Fig. 3c). To find the time $t^*$ when a probe starts, we first compute a spectrogram of the clipped signal with a hamming window of size $W = 512$ and overlap ratio $O = 0.5$. We then adopt template matching to the spectrogram as,

$$t^* = \underset{t}{\mathrm{argmax}} \sum_{i=1}^{M} F(t+i, S(i))$$

where $M = \frac{N}{WO}$ is the number of frames in one second spectrogram, $N = 48$ kHz is sound sampling rate, $F(i,j)$ represents absolute value of the spectrogram at $i$-th time frame and $j$-th frequency coefficient, and $S(i)$ is a template function defined as follows. For the single-frequency and multiple frequency probes, we use $S(i) = 5000$. For the linear sweep signal, $S(i)$ is defined as

$$S(i) = f_0 + \frac{f_1 - f_0}{M} i,$$

and for exponential sweep signal, we use

$$S(i) = f_0 \exp\left(\log\left(\frac{f_1}{f_0}\right) \frac{i}{M}\right)$$

where $f_0 = 100$ Hz and $f_1 = 10000$ Hz are the frequencies at the beginning and at the end of the probe, respectively.

Next, we extract a feature vector from the clipped signal. To analyze the reflected signal in the frequency domain, we test two types of features, such as spectrum (Fig. 3d) and Mel-Frequency-Cepstrum-Coefficients (MFCCs, Fig. 3e). For computing the spectrum, we applied a fast Fourier transform with hammin window to the one second clipped signal. For computing MFCCs, we used the window size of 1024 and the Mel-bands number of 20. The feature extraction process above was implemented by using Python.

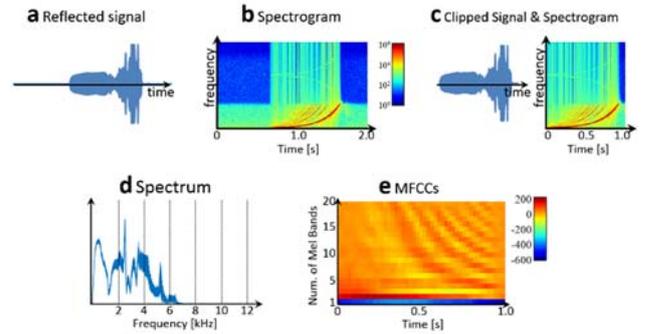

Fig 3. Feature vector extraction. Given a raw reflected signal (a), we compute its spectrogram (b), detect the time when each probe starts and cutout one second signal corresponding to a probe (c). We then compute spectrum (d) and MFCCs of the clipped signal (e).

### 4. Results and Discussion

To confirm the accuracy of our estimation technique, we performed experiments with tomatoes and mandarin oranges. After explaining the data collection process in detail, we will discuss the results of cross validation.

### 4.1 Data collection

We collected tomatoes and mandarin oranges and measured their firmness and reflected signals. In this study, it is important to collect samples with similar conditions at harvest. We therefore, built a dedicated vinyl house, cultivated tomatoes and mandarin oranges in it (Fig 4). Specifically, farmers managed the temperature, humidity, and moisture of the soil and cultivated tomatoes and mandarin oranges the same as they would usually. When harvesting, the famers checked the size, surface color, and stem color of the tomatoes and mandarin oranges to align the condition. All collected samples satisfied the sales standard of Japan Agricultural Cooperatives.

We harvested 162 tomatoes at the same time, divided them into 18 groups, and stored them in a cold (8 - 9 °C) and dark place. We measured the firmness and reflected signals of one group (9 tomatoes) every two days and recorded the storage time (day) of the group. To obtain large data sets, we measured the reflected signals and firmness at four different points on the tomatoes as in Fig 5a. In summary, we collected 9 (tomatoes) × 18 (groups) × 4 (points) = 648 sound files and 648 firmness values.

We harvested 180 mandarin oranges at one time, divide them in 20 groups and stored them in cold and dark place. Similar to the case of tomatoes, we measured the firmness and reflected signals of one group (9 mandarin oranges) every 3 days. We began the measurement on the 3rd day. The 6th, 13th and 17th groups could not be measured due to experimental circumstances. In summary, we collected 9 (mandarin oranges) × 17 (groups) × 4 (points) = 612 sound files and 612 firmness values.

The firmness was measured by a digital fruits firmness tester, *LUTORON FR-5120*, with a tip of 3 mm diameter. We push the tip of the tester to a sample and record a penetration pressure (Fig 5b). For the tomatoes, we peeled the thin skin before taking measurements to ignore counterforce caused by the skin (Fig 5c).

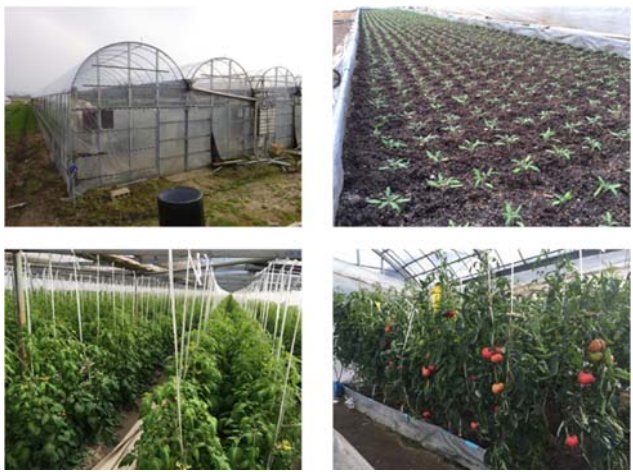

Fig 4. Environment for cultivating tomatoes and mandarin oranges.

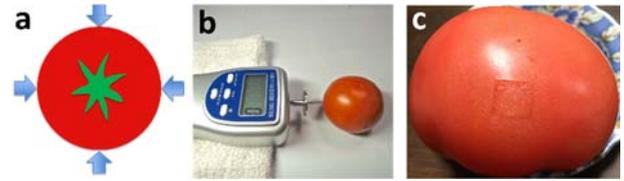

Fig 5. Data collection. We measured firmness and reflected signals of a sample at four different points (a). The firmness was measured by using a standard fruit firmness tester (b). For tomato samples, we peeled the thin skin to measure the firmness (c).

### 4.2 Cross Validation

To evaluate the accuracy of our acoustic probing, we performed $k$-fold cross validation ($k = 3$). We divided the collected data set into three groups and used two groups as a training set and the others for the test set. We then measured the accuracy of the regressor trained with the training set by using the test set. We repeated this process three times such that each of the three groups was used as a test set. Notice that, during data collection, we captured four reflected signals from one sample. When dividing the data into three groups, we avoided the condition where reflected signals comes from the same sample are classified into the both the training set and test set.

As mentioned, we tested four probe signals (i.e., single-frequency, multiple-frequency, linear-sweep, and exponential-sweep), two feature vectors (i.e., spectrum and MFCCs), and two regressors (i.e., SVR and GBR). Figs 6 and 7 summarize the average absolute errors of the estimation results in all conditions for tomato and mandarin orange, respectively. For the storage time, the minimum values of the average estimation errors were 0.89 days for tomatoes and 1.67 days for mandarin oranges. For the firmness, the minimum errors were 9.47 $g/mm^2$ for tomatoes and 15.67 $g/mm^2$ for mandarin oranges.

We found that the linear-sweep and exponential-sweep signals similarly provided highly accurate estimation comparing to the single-frequency and multiple-frequencies. This indicates that it is important for providing acoustic signals with various frequencies. Also, we did not find an apparent difference between the linear- and exponential-sweep signals. For the two feature vectors, spectrum and MFCCs, the spectrum achieved higher accuracy than MFCCs did in almost all conditions. This reason, we think, is that the number of Mel-bands was too low (we used 20) and important features were compressed. Note that a larger number of Mel-bands requires a larger memory size. Finding the best number of Mel-bands in terms of memory consumption and estimation accuracy remains as our future work. We also did not find an apparent difference between SVR and GBR.

Comparing the results of tomatoes and mandarin oranges, the accuracy of mandarin oranges was lower than that of tomatoes. We think this is caused by the anisotropic internal structure of mandarin oranges. A mandarin orange usually contains 10 or more segments. With this structure, the distance from a surface to segments differs depending on measurement points on a surface. This causes unstable measurements; reflected signal and firmness values may differ depending on points on a surface. In the future, we would like to solve this issue by developing a device that measures reflected signals at multiple points simultaneously and uses their average.

For more detailed analysis, we show the charts of all estimation results with the spectrum feature and SVR for tomato in Fig. 8 and those for mandarin oranges in Fig. 9. In each chart, the

### Average errors for Tomato data set

| SVR/GBR | Probe | Feature | Storage time | Firmness |
|---|---|---|---|---|
| SVR | Single Freq. | Spectrum | 4.58 | 16.37 |
| | | MFCC | 9.53 | 22.71 |
| | Multi. Freq. | Spectrum | 2.69 | 12.35 |
| | | MFCC | 4.90 | 22.50 |
| | Linear Sweep | Spectrum | 0.94 | 11.57 |
| | | MFCC | 2.85 | 18.62 |
| | Exp. Sweep | Spectrum | 1.08 | 10.06 |
| | | MFCC | 2.94 | 16.88 |
| GBR | Single freq. | Spectrum | 4.19 | 11.69 |
| | | MFCC | 6.49 | 12.90 |
| | Multi. freq. | Spectrum | 1.95 | 10.26 |
| | | MFCC | 3.63 | 11.01 |
| | linear Sweep | Spectrum | **0.89** | 9.51 |
| | | MFCC | 2.69 | 10.38 |
| | Exp. Sweep | Spectrum | 0.91 | **9.47** |
| | | MFCC | 3.01 | 10.69 |

Fig 6. Cross validation for the tomato data set. We summarize the average errors of storage time [days] and firmness [g/mm$^2$] with different regressors, probes, and features.

### Average errors for Mandarin Orange data set

| SVR/GBR | Probe | Features | Storage time | Firmness |
|---|---|---|---|---|
| SVR | Single Freq. | Spectrum | 11.87 | 21.70 |
| | | MFCC | 16.34 | 30.22 |
| | Multi. Freq. | Spectrum | 6.03 | 18.84 |
| | | MFCC | 12.41 | 32.44 |
| | Linear Sweep | Spectrum | 1.81 | 21.90 |
| | | MFCC | 5.59 | 27.86 |
| | Exp. Sweep | Spectrum | **1.67** | 17.41 |
| | | MFCC | 5.18 | 28.42 |
| GBR | Single Freq. | Spectrum | 10.16 | 16.87 |
| | | MFCC | 11.14 | 17.59 |
| | Multi. Freq. | Spectrum | 4.36 | 17.23 |
| | | MFCC | 7.94 | 16.78 |
| | Linear Sweep | Spectrum | 2.85 | **15.67** |
| | | MFCC | 5.00 | 16.66 |
| | Exp. Sweep | Spectrum | 2.12 | 15.69 |
| | | MFCC | 5.07 | 17.17 |

Fig 7. Cross validation for mandarin orange data set. We summarize the average errors of storage time [days] and firmness [g/mm$^2$] with different regressors, probes, and features.

horizontal and vertical axes represent the measured (i.e., ground truth) and the estimated values, respectively. In the charts for storage time estimation with the linear sweep signal (Figs 8c and 9c) and the exponential sweep signal (Figs 8d and 9d), plots are closely aligned to the diagonal line, which indicates that highly accurate estimations were achieved. However, in the charts for firmness estimation with linear sweep signal for tomatoes (Fig 8g), plots are widely distributed. Furthermore, in the chart for mandarin oranges (Fig 9g), we could not find a correlation between the measurement and estimation. In summary, our technique achieves highly accurate estimation for storage time for tomatoes and mandarin oranges and moderately accurate estimation for firmness of tomatoes, but fails to estimate the firmness of mandarin oranges.

## 5. Conclusions and Future work

In this paper, we have presented an acoustic probing technique for estimating storage time and firmness of fruits. We provide an acoustic signal by using a small microphone that gently contacts to a sample and capture reflected signals. We then extract a feature vector from the captured signal to train a regressor. Since our technique only requires a small speaker and a microphone, it is possible to make a measurement device small and produce it at a low cost. To evaluate the accuracy of our technique, we conducted experiments by using tomatoes and mandarin oranges. As results, it achieved highly accurate estimation for the storage time for the both fruits; however, we found limitation in firmness estimation of mandarin orange. By testing four different probes, we found that the selection of acoustic probe is important for accurate estimation and probes containing various frequencies, i.e., sweep signals, achieves highly accurate estimation.

As we discussed, our technique was limited in estimating firmness of mandarin orange and this limitation seems to be caused by the anisotropic internal structures of this fruit. This problem will be solved by performing measurement at multiple points. Another limitation of our technique is to deal with fruits with thick and hard peel, such as pineapples or watermelons, because the acoustic signal generated by a small speaker (earphone) does not pass through thick peels of such fruits. To solve this problem, in the future, we would like to adopt different probes, such as vibration. Other future work includes ways to adopt our technique to different fruits with thin peels and to implement a measurement device by using a smartphone to illustrate the practical usefulness of our acoustic probing.

### Acknowledgements

This work was supported by Japan Society for the Promotion of Science Grants-in-Aid for Scientific Research (15H05924) to TI and AH and by Information-Technology Promotion Agency (IPA) Mito Software Project to HK.

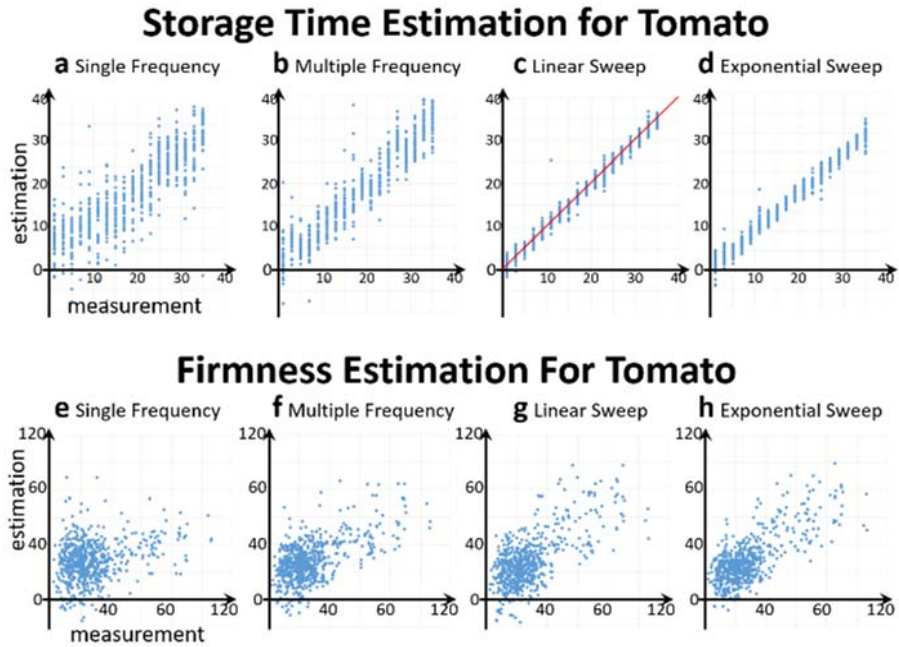

Fig 8. Detailed estimation results for tomato with the spectrum feature and SVR. In each chart, the horizontal and vertical axes represent the measured and estimated values, respectively. We have plotted 648 estimated data points in each chart.

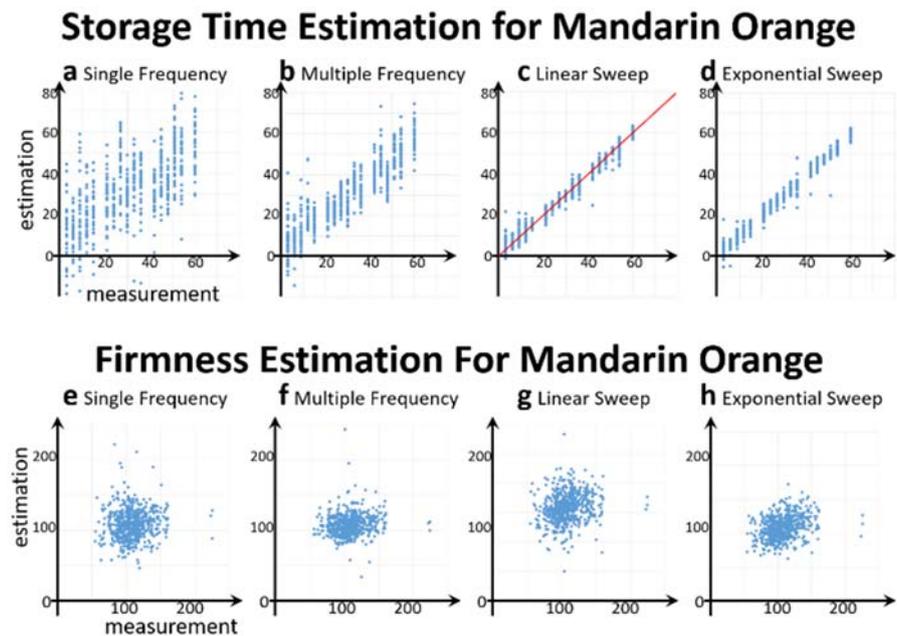

Fig 9. Detailed estimation results for mandarin oranges with the spectrum feature and SVR. In each chart, the horizontal and vertical axis represent the measured and estimated values, respectively. We plot 612 estimated data points in each chart.